# Understanding the development of interest and self-efficacy in active-learning undergraduate physics courses

Remy Dou [a,b], Eric Brewe [c], Geoff Potvin [b,d], Justyna P. Zwolak [e] and Zahra Hazari [a,b,d]

[a]Department of Teaching and Learning, Florida International University, Miami, FL, USA; [b]STEM Transformation Institute, Florida International University, Miami, FL, USA; [c]Department of Physics, Drexel University, Philadelphia, PA, USA; [d]Department of Physics, Florida International University, Miami, FL, USA; [e]Joint Center for Quantum Information and Computer Science, University of Maryland, College Park, MD, USA

**ABSTRACT**

Modeling Instruction (MI), an active-learning introductory physics curriculum, has been shown to improve student academic success. Peer-to-peer interactions play a salient role in the MI classroom. Their impact on student interest and self-efficacy – preeminent constructs of various career theories – has not been thoroughly explored. Our examination of three undergraduate MI courses (N = 221) revealed a decrease in students' physics self-efficacy, physics interest, and general science interest. We found a positive link from physics interest to self-efficacy, and a negative relationship between science interest and self-efficacy. We tested structural equation models confirming that student interactions make positive contributions to self-efficacy. This study frames students' classroom interactions within broader career theory frameworks and suggests nuanced considerations regarding interest and self-efficacy constructs in the context of undergraduate active-learning science courses.

## Introduction

A national imperative exists to motivate and empower undergraduate students in the pursuit of physics degrees, which includes a push for the implementation of effective active-learning modalities (National Research Council [NRC], 2013). A variety of reasons have spurred this call, including the lack of representation from members of minority groups in physics and also lack of adequate science teacher preparation in physics. With regard to career choice, persistence, and even teacher quality, the contribution of domain-specific self-efficacy and interest has been well-examined (Bandura, 1993, 1997; Holzberger, Philipp, & Kunter, 2013; Klassen, Tze, Betts, & Gordon, 2011; Schunk, 2012). In particular, these constructs have been shown to predict career choice behaviour more so than academic achievement or personality-based theories (Bandura, Barbaranelli, Caprara, & Pastorelli, 2001; Kjærnsli & Lie, 2011; Lent, Brown, & Larkin, 1987). Lent, Brown, and Hackett's (1994) Social Cognitive Career Theory (SCCT) affirms the mediating role that self-efficacy plays between learning experiences and interest. In turn student interest can have an effect on students' goals and goal accomplishments – like completing a physics major. Carlone and Johnson (2007) make the case that 'performance competence,' a construct that shares some similarities with self-efficacy, influences students' interest and recognition in a particular field, which can then contribute to how students identify with that field. This field-specific identity framework has been shown to predict career choices of physicists and engineers (Godwin, Potvin, Hazari, & Lock, 2016; Hazari, Sonnert, Sadler, & Shanahan, 2010). Given our current national focus on imple-menting student-centered, active-learning curricula, as a way to address inequalities in our system, how these pedagogies impact career-shaping constructs like self-efficacy and interest deserves attention, especially in light of the social context of active-learning curricula (National Academy of Sciences, 2016; President's Council of Advisors on Science and Technology, 2010). The current work represents an effort to examine the relationship between student interactions that take place during active-learning introductory physics courses and changes in students' physics self-self-efficacy, physics interest, and broader science interest.

CONTACT Remy Dou  redou@fiu.edu  Department of Teaching and Learning, Florida International University, 11200 SW 8TH ST, Miami, FL 33199, USA; STEM Transformation Institute, Florida International University, 11200 SW 8TH ST, Miami, FL 33199, USA

### Similarities across two prominent career theories

The focus of this study on 'interest' and 'self-efficacy' is driven by their predictive value with regard to career-choice. Two prominent career theories, the SCCT and the identity framework alluded to above include these constructs in their models. Both maintain direct links from self-efficacy to constructs that mediate career choice (i.e. choice goals and identity, respectively). Both also maintain indirect links mediated by interest between self-efficacy and these career choice constructs.

The SCCT was developed by Lent et al. (1994) as a model of career choice. Research in a variety of science education contexts has verified their work (e.g. Dickinson, 2007; Lent, Lopez, Lopez, & Sheu, 2008; Navarro, Flores, & Worthington, 2007; Smith & Fouad, 1999). Meta-analyses have confirmed model agreement across these contexts (Brown et al., 2008; Sheu et al., 2010). The SCCT suggests that learning experiences mitigated by person-inputs (e.g. gender, race/ethnicity) contribute to a person's self-efficacy in a particular domain (e.g. physics). Bandura (1993) defines this domain-specific self-efficacy as individuals' beliefs about their ability to complete particular tasks within that domain. The SCCT also proposes that learning experiences contribute to individuals' 'outcome expectations' around certain tasks and domain-specific work. Self-efficacy and outcome expectations in turn predict students' career interests, goals, and related behaviour. The SCCT has been corroborated with various populations, including young children, college students, and adults, as well as people pertaining to minority groups (Nauta, Kahn, Angell, & Cantarelli, 2002).

Findings from the work of Carlone and Johnson (2007), Gee (2001) and Shanahan (2007, 2008) have shown that identity (i.e. a person's sense of self) is also a significant and meaningful predictor of career choice. For example, Hazari et al. (2010) show that individuals who see themselves as physicists (i.e. have high physics identity) are more likely to pursue careers in physics. Their work points to three specific constructs as predictors of identity: performance/competence, interest, and recognition. This identity frame work of vocation posits that performance/competence predicts individuals' interest in a particular field of work and sense of recognition from members who participate in that field. Interest and recognition in turn predict individuals' sense of identity in a particular domain.

### Career choice constructs: self-efficacy and interest

The constructs of self-efficacy and interest have been found to predict students' academic performance (e.g. grades) and career choices (Bandura et al., 2001; Riegle-Crumb, Moore, & Ramos-Wada, 2011; Shen, 2002). In physics, the direct relationship between grades and self-efficacy has been seen in contexts that differ widely, including private institutions in New England and public universities in South Florida (Lynch, 2010; Sawtelle, Brewe, Goertzen, & Kramer, 2012). Academic settings outside of the United States support similar findings (e.g. Lindstrøm & Sharma, 2011). Links between interest and career persistence have also been established. For example, data from a random national sample of students from 34 different higher education institutions showed a strong relationship between physics interest in high school and intent to pursue a STEM career in college (Sadler, Sonnert, Hazari, & Tai, 2012).

Physics interest and self-efficacy, though distinct constructs with regard to how they explain variance in career-choice (Donnay & Borgen, 1999; Smith & Fouad, 1999), are often found to correlate with values ranging from 0.20 to 0.70 (Bieschke, Bishop, & Garcia, 1996; Lenox & Subich, 1994; Lopez, Lent, Brown, & Gore, 1997; Tang, Fouad, & Smith, 1999). The relationship between these two constructs is understood as reciprocal, one informing the other. Nevertheless, Nauta et al. (2002) point out that models like the SCCT tend to emphasise the preeminence of self-efficacy as a primary contributor to interest development (e.g. Bandura, 2001; Godwin et al., 2016; Lent et al., 2008; Sheu et al., 2010; Smith & Fouad, 1999). In fact, the inverse can also occur such that initial interest in a field drives self-efficacy development particularly in contexts where participants are engaging in novel experiences (Bandura, 1997; Lent et al., 1994; Nauta et al., 2002).

## Modeling instruction

With the expansion of physics education research, introductory physics courses have come a long way from the from traditional lecture-based teaching styles (e.g. Etkina & Van Heuvelen, 2007; Finkelstein & Pollock, 2005). Many if not all of these differ from tra-ditional instruction in that they solicit a greater number of student-student interactions. This student-centric approach is at the core of much of the reform in undergraduate physics education (NRC, 2013). Given the propagation of student-centric strategies implemented in active-learning courses, their impact on student learning merits attention, especially in light of the relationship between peer interactions and the development of constructs like self-efficacy (see below).

Of the various active-learning curricula in introductory physics, Modeling Instruction (MI) has shown positive results with regard to students' academic performance and atti-tudes toward science in both high school and undergraduate contexts (Brewe, 2008; Hes-tenes, 1987). MI Introductory Physics I and II with Calculus courses (referred to as MI from here on out) were created by Hestenes (1987) and further developed for college-level introductory physics by Desbien (2002) and Brewe (2008). A prominent feature of the MI curriculum is its application of student discourse as a tool for knowledge construction, particularly around a few, key disciplinary concepts. Students work in small and large groups on carefully designed, inquiry-based activities meant to stimulate discussion. For example, students may find themselves walking in front of motion detectors as part of the constant acceleration unit. They are prompted to think about a general model that identifies the major concepts (e.g. position, displacement, velocity), which they must then develop multiple representations for using motion diagrams and kinematic graphs, for example. Later, small groups come together into what is called a 'board meeting,' where they use portable white boards to share and discuss their representations with other groups. As a result of these activities, students can often be seen initiating conversations with their tablemates or collaborating with peers at other parts of the room in order to reach consensus on course-related tasks.

Students who enroll in MI pass at greater rates than students taking equivalent lecture-based courses and score higher on concept inventories (Brewe et al., 2010), suggesting that students gain a better understanding of physics. These findings align with Freeman et al.'s (2014) meta-analysis that affirms the benefit of active-learning science courses with regard to student achievement on exams and concept inventories. Yet, physics students in both traditional and active-learning physics classrooms often see a decrease in their physics self-efficacy (Nissen & Shemwell, 2016) – a decline that stands contrary to the positive aca-demic outcomes. In Dou et al. (2016) the overall decrease in physics self-efficacy was observed regardless of student gender or ethnicity, and suggested the presence of structures in the active-learning curricula that may negatively impact the factors that motivate students to pursue careers in STEM. These structures may include, among many, students' social interactions in the classroom. In this particular study a subset of students who received attention from academically popular peers (i.e. perceived by others as meaningful academic resources) were more likely to see increases in their individual self-efficacy scores.

### *The relationship between classroom interactions and physics self-efficacy formation in MI classrooms*

Qualitative analyses of student participation in MI courses revealed the existence of opportunities for students to build their physics self-efficacy. These self-efficacy experience opportunities (SEOs) are aligned with one of the four sources of self-efficacy (i.e. mastery experiences, vicarious learning, verbal persuasion, physiological states) that have the potential to impact its development (Sawtelle et al., 2012). In their research, Sawtelle et al. (2012) highlight students discussing a position versus time graph two as an example of an SEO. The students discussed the importance of having a reference point and mutual agreement on what the reference point was in the problem they were solving served as a potential mastery experience. The authors concluded that the abundant presence of these types of experiences resulted from the discourse-eliciting group activities that are part of the MI curriculum. Similarly, Bandura (1977) and Pajares (1997) posit an inherent relationship between self-efficacy development and social inter-actions, specifically citing the role of vicarious learning and verbal persuasion experiences. Using social network analysis (SNA) to quantify student interactions, Dou et al. (2016) found a correlation between students' position in their classroom social network and improved self-efficacy development. Similarly, classroom network measures, like PageRank centrality (see Methods) have exhibited positive relationships with gains on the Force-Motion Concept Evaluation (FMCE) – a well-established introductory physics concept inventory (Thornton & Sokoloff, 1998; Williams, Brewe, Zwolak, & Dou, 2015). This, too, aligns with past studies showing better student under-standing when constructing knowledge in group settings (Alexopoulou & Driver, 1996; Stump, Hilpert, Husman, Chung, & Kim, 2011). These relationships between inter-actions, self-efficacy, and learning frames our examination of peer-to-peer relationships in active-learning physics courses and their association with the development of self-efficacy and interest.

## Purpose

We sought to further situate the contribution of students' classroom interactions in the larger context of physics self-efficacy and physics interest development. Both SCCT and identity framework posit that self-efficacy and interest directly and indirectly influence the constructs antecedent to STEM career choice, yet the role of classroom interactions on the evolution of these two former constructs is only partially understood (Dou et al., 2016; Hazari et al., 2010; Lent et al., 1994). By measuring classroom interactions and testing various model pathways between these variables (i.e. interaction patterns, self-efficacy, interest), we set out to further understand the directional influence of classroom interactions. Given the reciprocal connection between self-efficacy and interest, we also tested models with varying paths with the goal of answering the following research questions (see Figure 1):

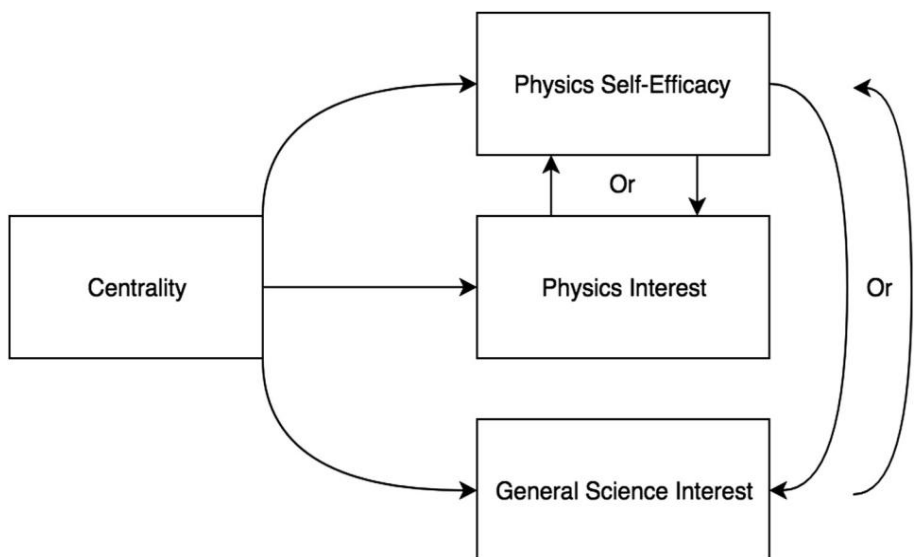

Figure 1. General model indicating hypothesised and alternative relationships between centrality, self-efficacy, and interest.

(1) Do classroom interactions as measured by social network centrality directly influence students' science and physics interest in MI courses?
(2) Do classroom interactions as measured by social network centrality indirectly influence students' science and physics interest via their physics self-efficacy in MI courses?
(3) Do reciprocal relationships exist between self-efficacy and interest in an active-learn-ing introductory physics course?

## Methods

### Data

Data for this study came from three MI courses taught in the fall 2014 and fall 2015 semesters at a large Hispanic Serving Institution (HSI) in South Florida. While the majority of students at this institution register for traditional sections of Introductory Physics I with Calculus taught in auditorium-style settings, a few sections of MI become available to all students, permitting their schedule and requisites complement the course. Students sign up on a first-come-first-served basis. In 2014 active-learning science classrooms with movable chairs and tables became available with a capacity of approximately 80 students. Only one section of MI was offered in fall 2014 and in fall 2015 two sections were offered. The same instructor who taught in fall 2014 also taught in fall 2015. An additional instructor taught the second MI section offered in fall 2015. Both instructors had experience with active-learning teaching strategies geared at intro-ductory physics students and both received support from faculty in the physics depart-ment, as well as Teaching Assistants and Learning Assistants, in the implementation of the MI curriculum.

The majority of students from these three sections (N = 221) identified as Hispanic (i.e. 71%) and the remaining students classified themselves as Asian, White, or Black (see Table 1). They matriculation survey from which this data was retrieved did not dis-tinguish between 'race' and 'ethnicity'. The majority of these students were male (i.e. 56%), while 43% identified as female. Demographic and/or gender information for six students were not available. Of the 35 majors represented in these courses, only about 1% of stu-dents had elected to pursue a bachelor's degree in physics at the time they took the course. The majors with greatest representation included Biology (i.e. 24%), Mechanical Engineering (i.e. 12%), and Computer Science (i.e. 10%). Nearly all remaining students were pursuing majors in a STEM domain. At this institution, introductory physics is a requisite for most STEM degrees.

Table 1. Participant characteristics.

| Demographics | |
|---|---|
| Gender | Female = 43% |
| | Male = 56% |
| Ethnicity (population greater than 1%) | Asian = 6.3% |
| | Black = 10% |
| | Hispanic = 71% |
| | Two or more = 1.4% |
| Majors (representation greater than 1%) | Biology, BS = 24% |
| | Biomedical Eng., BS = 7.2% |
| | Chemistry, BA = 5.0% |
| | Chemistry, BS = 3.2% |
| | Civil Eng., BS = 4.5% |
| | Computer Eng., BS = 4.5% |
| | Computer Science, BS = 10% |
| | Dual Enrollment, high school = 3.6% |
| | Dual Major = 2.7% |
| | Electrical Eng., BS = 2.7% |
| | Environmental Eng., BS = 1.4% |
| | Experimental Psych. = 3.6% |
| | Mechanical Eng., BS = 12% |
| | Psychology, BA = 2.7% |

### In-class social network data

Developed by sociologists in the 1930s, SNA allows researchers to quantify interactions between people and examine hidden structures found in their relationships (Scott, 1988). SNA has revealed complexity in the social and academic networks of physics students, changes in study-group networks and their relationship with biology exam performance, and how students from varying STEM majors participate in informal student networks (Forsman, Moll, & Linder, 2014; Grunspan, Wiggins, & Goodreau, 2014; Sadler et al., 2012). Here we were particularly interested in measuring student 'centrality.' Centrality refers to a specific node's (i.e. individual) relational position in a network and can come in many forms (Scott & Carrington, 2011). For example, in the case of a classroom network, *directed degree* centrality quantifies the total number of incoming and outgoing interactions experienced by a particular student, while *PageRank* centrality ranks students according to the popularity of the peers with whom they interact using an algorithm initially developed by Brin and Page (1998) for the Google search engine. The current study built on our use of these tools in undergraduate physics contexts.

We collected student interaction information by surveying students five times during the semester, asking them the following:

> Please choose from the presented list people from your physics class that you had a meaningful interaction with *in class* this week, even if you were not the main person speaking or contributing. You may include names of students outside of the group you usually work with.

Beneath the prompt students were afforded blank space divided into three columns where they could indicate both who they worked with, as well as how often they worked with each person (see Supplementary Files). Students were also asked about the individuals they worked with on physics-related material *outside* of class. For the purpose of this study we did not address students' out-of-class interactions. After the fall 2014 semester, we began providing students with randomly ordered lists of their peers' and instructors' names for the sake of efficiency and to avoid name confusion.

In general, we administered social network surveys on the last day of class during a particular data collection week. Weeks were chosen ahead of time and were distributed throughout the semester. Five network data collection events took place: one at the beginning of the semester, one at the end, and three in between. We strove to ensure that data collection happened on weeks when students participated in active-learning activities that required both large and small group cooperation and discourse. We achieved this setting for the first four data collection events each term. Nevertheless, in the fall of 2014 the last survey collection of the semester took place on a week that became dedicated to final exam review and student participation in inquiry activities was not solicited. Response rates exhibited this contrast, coming in at greater than 75% percent for survey administrations during the first four collection events of a semester and at less than 50% on the last survey given in the fall 2014 semester. Given the sensitivity of networks to context, responses from the fifth survey administered that semester were dropped, and subsequently the fifth administration was also dropped for all other courses in order to maintain homogeneity of network data and avoid inflated network measures of students in fall 2015. However, even after performing separate analyses of fall 2014 data and fall 2015 data the outcomes agree with the results of the combined semesters' analyses presented in this paper. Student responses to each survey were used to generate edgelists, which contained interaction information about the student responding to the survey (i.e. source) and the students listed as responses to the social network queries (i.e. targets). We then used the resulting student inter-action matrix to calculate student centrality using the igraph package in R (Csárdi & Nepusz, 2006; R Core Team, 2015). To generate an overall measure of student inter-actions we compiled responses across the first four social network surveys administered during each course by pooling the edgelists.

Duplicate interactions between the same two individuals across the semester were weighed according to the number of occurrences. For example, if student A named student B as someone with whom he or she had a meaningful interaction with during the first week of the semester, that edge (i.e. link between two nodes) was given a weight of one. If student A named student B on one or more surveys administered later in the semester, the edge between A and B received a weight of plus one for each instance. We selected this approach in order to capture the overall structure and strength of stu-dents' networks across the semester. The final pooled edgelist was used to calculate par-ticipants' PageRank centrality. We concentrated on students' PageRank centrality for two primary reasons: (a) PageRank incorporates the number of incoming and outgoing interactions students experience, as well as the kinds of peers with whom they interact, and (b) PageRank centrality, unlike degree centrality, which is a measure more commonly used in education research, has been shown to be predictive of students' overall self-efficacy development (Dou et al., 2016).

PageRank values for an individual student '$i$' was calculated as depicted in equation 1, where $p(j)$ is the PageRank of student $j$, $k_{out}(j)$ is the number of outgoing interactions of student $j$, $n$ is the number of students, and $q$ is a damping factor set at 0.15, which accounts for the possibility of random, disconnected interactions between any two individuals in the classroom (Fortunato & Flammini, 2007):

$$p(i) = \frac{q}{n} + (1-q) \sum_{j:j \to i} \frac{p(j)}{k_{out}(j)} \qquad i = 1, 2, \ldots, n. \qquad (1)$$

In short, PageRank calculates the likelihood that a person travelling through the network of interactions would reach a particular individual (Csárdi & Nepusz, 2006). After calculating PageRank we removed instructors from the final network in order to focus on peer interactions.

*Physics self-efficacy*

Students completed the Sources of Self-Efficacy in Science Courses – Physics (SOSESC-P) survey at the beginning and end of the MI courses. We selected this survey because of its introductory physics context specificity – a characteristic of effective self-efficacy surveys (Pajares, 1997) – and to maintain research continuity with previous studies in active-learning introductory courses. Students indicated how much they agreed with a variety of statements using a 5-point Likert scale. Statements included the following: '*I will have difficulty with the exams/quizzes in this class*' and '*I will get positive feedback about my ability to recall physics ideas.*' We achieved an alpha reliability coefficient of .92 for the instrument. This 33-item survey was designed to both measure the sources of self-efficacy (i.e. mastery experiences, vicarious learning, verbal persuasion, physiological mechanisms) and act as a proxy for students' overall self-efficacy (Fencl & Scheel, 2005; Sawtelle et al., 2012). It is worth noting that prominent researchers (e.g. Usher & Pajares, 2008) have warned the field about combining items meant to measure the different sources of self-efficacy. With regard to the SOSESC-P, past studies have set precedence for reliably using this instrument to measure overall self-efficacy in introductory physics contexts (e.g. Fencl & Scheel, 2004; Trujillo & Tanner, 2014), and confirmatory factor analysis of its items has affirmed its use as a general self-efficacy measure (Dou, 2017). Nevertheless, we present this as a possible limitation of our study.

*General science and physics interest*

Interest in science and physics related content was collected at the beginning and end of the class using the Physics Identity Development (PID) survey (Potvin & Hazari, 2013). This questionnaire included items related to students' sense of identity as a physics person, their sense of performance competence, their general science interest, and their physics interest. In the current study, we were only concerned with students' responses to questions about general science interest (i.e. understanding natural phenomena, under-standing science in everyday life, explaining things with facts, telling others about science concepts, making scientific observations, wanting to know more science) and physics interest (i.e. mechanics, optics/waves, electromagnetism, relativity/modern physics, history and people in physics, current topics in physics). Indication of interest was demarcated on a seven-point anchored Likert scale ranging from 'No Interest' to 'Very Interested'. We achieved an alpha coefficient of reliability of .89 on the general science interest items and .88 on the subset of physics interest items. Confirmatory factor analysis affirmed the use of the aggregate of these items as a measure of students' overall science and physics interest, respectively (see 'Analysis & Results') (Table 2).

Table 2. Summary of student responses to surveys.

| Variables | |
| --- | --- |
| Centrality | PageRank, (M = 0.012, SD = 0.00222) |
| Physics Self-efficacy | pre SOSESC-P, (M = 91.5, SD = 10.9) |
| | post SOSESC-P, (M = 87.7, SD = 15.0) |
| Physics Interest | pre, (M = 25.2, SD = 6.80) |
| | post, (M = 23.0, SD = 7.95) |
| General Science Interest | pre, (M = 30.2, SD = 5.19) |
| | post, (M = 29.3, SD = 5.95) |

## Results

### Factor analyses and t-tests

Using confirmatory factor analyses we examined our measurement model to test how well our survey items measured physics self-efficacy and interest. Single factor loadings for physics self-efficacy items on the SOSESC-P ranged from 0.41 to 0.85 after removing eight items that fell below a 0.40 cutoff and running single imputation on our data to account for missingness (Stevens, 1992). All of the items on our physics interest survey fell within an acceptable single factor loading range of 0.51 to 0.88. Similarly, the items on general science interest presented factor loadings between 0.68 to 0.81. Allowing covariances among the constructs yielded a moderate fit for the measurement model ($\chi^2$ = 5795.2, df = 2261, $\chi^2$/df = 2.56, RMSEA = 0.075, SRMR = 0.078).

Prior to proceeding with testing our structural models, we performed dependent samples t-tests to determine whether student self-efficacy and interest changed throughout the semester. Mean physics self-efficacy scores decreased significantly from pre to post by 4.85 points. This indicated that students reported lower physics self-efficacy at the end of the course than they did at the beginning [$t(220) = 5.21, p < .001$, Cohen's $d = 0.35$]. This represented a 7.35% drop based on the overall range of student scores (i.e. 59–125). Simi-larly, students' physics interest decreased significantly by 2.23 points or 6.2% of the scale [$t(220) = 5.24, p < .001$, Cohen's $d = 0.35$]. A small, but significant negative trend of one-point was also observed in students' general science interest from pre to post [$t(220) = 2.84, p < .01$, Cohen's $d = 0.19$].

### Structural equation modeling

Our use of structural equation modeling (SEM) is predicated by its handling of multiple dependent variables and robustness of its incorporation of latent variables (Harlow, 2014). SEM can be generally described as a two-part process, which begins with a CFA of the latent variables (i.e. variables that cannot be directly observed). This analysis tests for the hypothesised number of latent factors measured by a set of items. In our study, we measured three latent variables (i.e. physics self-efficacy, physics interest, general science interest) using two different instruments (i.e. SOSESC-P and PID). The analysis of our CFA described in the previous section confirmed our hypotheses. The second part of an SEM allows researchers to test complex, multipath relationships between constructs where a plurality of both dependent and independent variables exist.

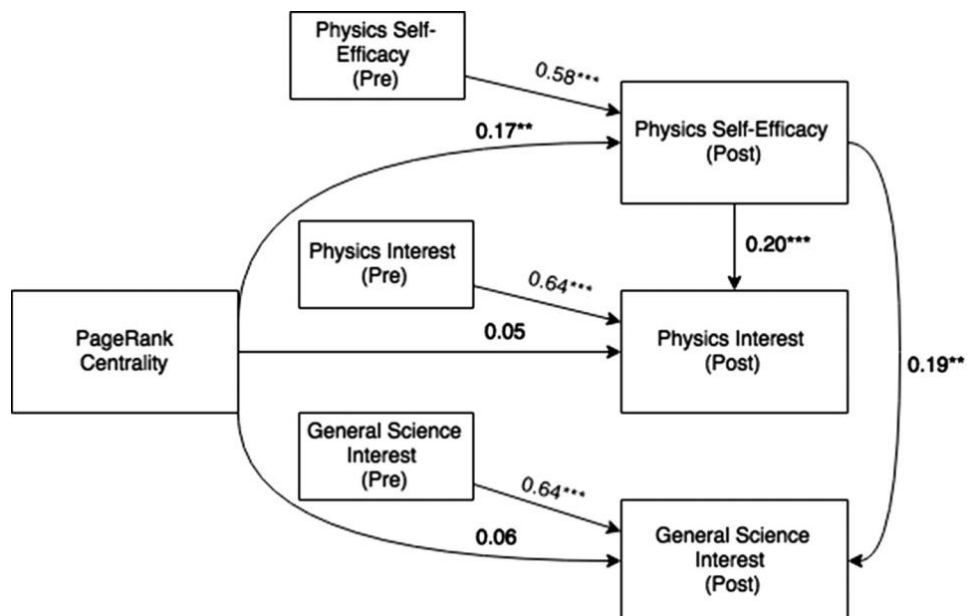

Figure 2. Results of Model A showing standardised path coefficients and significance (***$p < 0.001$, **$p < 0.01$). Item coefficients can be found in Table S1.

We tested our primary model (i.e. model A) with paths running directly from PageRank centrality to post physics self-efficacy (i.e. at the end of the course), post physics interest, and post general science interest, as well as a path from post physics self-efficacy to post physics interest (see Figure 2). In this model we also controlled for students' physics self-efficacy and interests at the beginning of the course (i.e. pre). The path leading from PageRank directly to post physics self-efficacy was significant ($\beta = 0.17$, $p < .01$), as well as the path from post physics self-efficacy to post physics interest ($\beta = 0.20$, $p < .001$) and general science interest ($\beta = 0.19$, $p < .001$). The paths from PageRank to post physics and general science interest were not significant. Item path coefficients are included in Table S1 (see Supplementary Files). This model had moderately acceptable fit measures [$\chi^2(2333) = 5944.5$, $p < 0.001$; $\chi^2/df = 2.55$, RMSEA = 0.08, SRMR = 0.08, BIC = 36210], which may suggest weak relationships between some of our variables, leaving more variance to be explained (please see Discussion).

Because of the reciprocal relationship sometimes exhibited between self-efficacy and interest, we tested a model similar to model A, reversing the path from post physics self-efficacy to post physics and general science interest (i.e. model B; see Figure 3). This model, too, revealed significant paths from PageRank centrality to post physics self-efficacy ($\beta = 0.15$, $p < .05$) and from post physics interest to post physics self-efficacy ($\beta = 0.28$, $p < .001$). Models A and B shared similar fit measures [$\chi^2(2334) = 5968.5$, $p < 0.001$; $\chi^2/df = 2.56$, RMSEA = 0.075, SRMR = 0.082, BIC = 35,535].

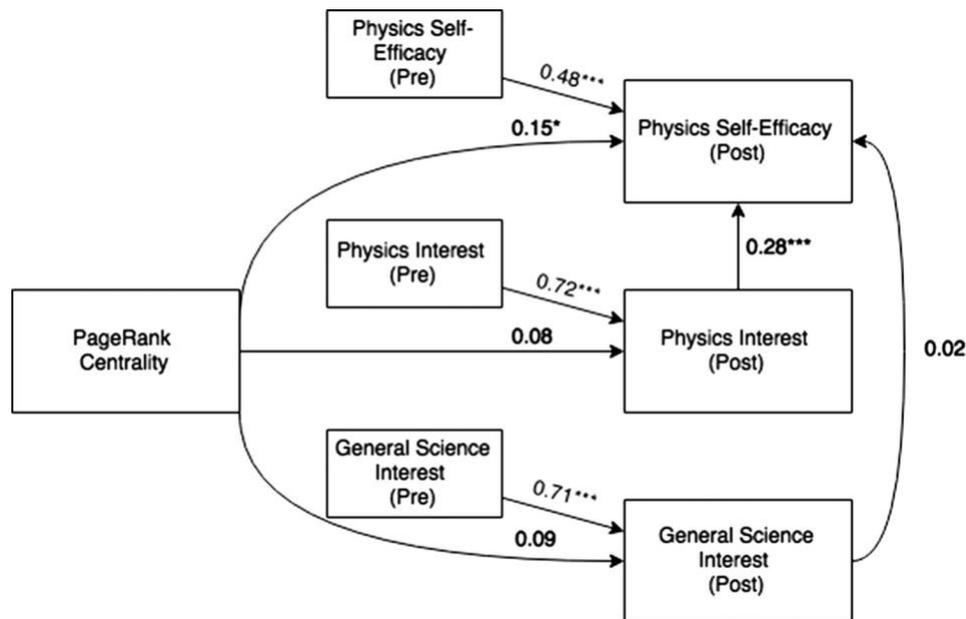

Figure 3 Results of Model B showing standardised path coefficients and significance (***$p < 0.001$, **$p < 0.01$). Item coefficients can be found in Table S1.

In order to better grasp the directionality between physics self-efficacy and interest we ran a third model (i.e. model C), taking advantage of the longitudinal nature of our data to determine whether students' self-efficacy contributed to their interest or vice versa. We included paths from students physics self-efficacy at the beginning of the course to their physics and general science interest at the end of the course. Similarly, we included paths from interest at the beginning of the course to physics self-efficacy at the end of the course (see Figure 4). The remaining paths mirrored those of models A and B. Paths common to models A and B exhibited similar significance. The paths from students' physics self-efficacy at the start of the course to their physics and general science interest at the end of the course were not significant, while the paths from physics and general science interest at the start of course to post physics self-efficacy were statistically significant ($\beta = 0.46$, $p = 0.01$; $\beta = -0.46$, $p = 0.01$). It is worth noting that the correlation between general science interest at the beginning of the course and physics self-efficacy at the end of the course was negative. This model yielded acceptable fit measures [$\chi^2(2329) = 5904.5$, $p < 0.001$; $\chi^2/df = 2.54$, RMSEA = 0.075, SRMR = 0.077, BIC = 36,192]. The results given by the test of this model indicated that students' interest in course content contributed to their self-efficacy development and not the other way around as we hypothesised.

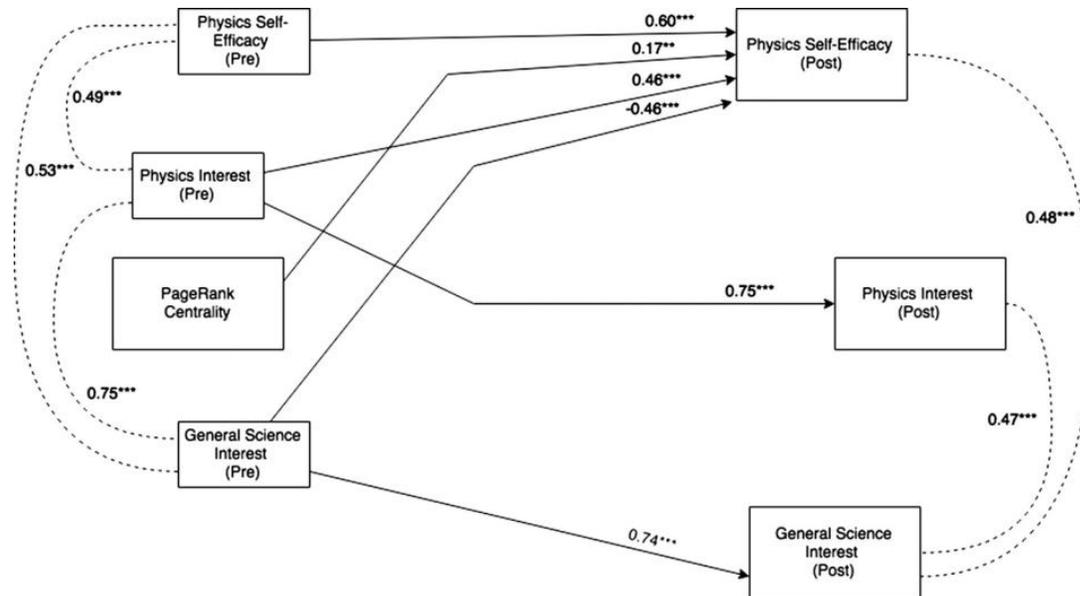

Figure 4 Results of Model C showing standardised path coefficients and significance (***$p < 0.001$, **$p < 0.01$). Only significant paths have been included. Item coefficients can be found in Table S1 (see Appendix).

Given the models' similarities in fit measures and Bayesian information criterion (BIC) values, we chose to compare them using log likelihood ratio tests. The additional parameters in model C explain significantly more of the variance in our data than model A and B [$\chi^2(4) = 38.0$, $p < 0.001$ and $\chi^2(5) = 62.0$, $p < 0.001$, respectively]. Model A also outperformed model B [$\chi^2(1) = 24.0$, p < 0.001], further supporting a latent variable directionality going from interest to self-efficacy. All three models (i.e. A, B, and C, respectively) explained significantly more variance than similar models without the PageRank centrality variable [$\chi^2(3) = 2084.4$, $p < 0.001$; $\chi^2(3) = 2086.7$, $p < 0.001$; $\chi^2(3) = 2081.8$, $p < 0.001$].

## Discussion

The results of our analyses offer different starting points for interpretation. The most straightforward of these involve the outcomes of our dependent samples t-tests that indicated a decrease in students' physics self-efficacy and physics interest. The documented benefit of active-learning curricula with regard to student learning stands seemingly contrary to these drops, which have also been observed in a variety of introductory physics contexts (Nissen & Shemwell, 2016). While on the one hand past research shows MI helps students develop a better understanding of physics content and experience smaller drops in self-efficacy than those in lecture sections (Brewe et al., 2010), participants in this study left the class feeling less confident about their ability to complete physics-related tasks and less interested in physics subjects, as well as exhibited a slight drop in general science interest. We might attribute this development directly to the curriculum, but prior research suggests two other possibilities: (a) re-calibration of students' perspectives and (b) reaffirmation of students' non-physics career intentions. The former explanation aligns with research indicating over-confidence or unrealistic expectations that incoming students in introductory courses hold (Boekaerts & Rozendaal, 2010; Lindstrøm & Sharma, 2011). Students' reassessment of their abilities in contrast to course experiences could lead to more accurate, and possibly lower, judgments of their actual abilities at the end of the semester. Moreover, Redish, Saul, and Steinberg (2000) have found that students place less effort toward the end of an introductory physics semester, a fact that aligns with the link between self-efficacy and effort expended (Bandura, 1997). The dearth of students intending to major in physics also participants came in with less interest in physics in comparison to science fields (Hazari et al., 2010; Lent et al., 1994). Their experiences in the course may reinforce their lack of interest in physics content and is reflected in the negative relationship between general science interest and physics self-efficacy at the end of the course.

We also discovered that students' PageRank centrality significantly and positively predicted students' physics self-efficacy at the end of the course even when controlling for their self-efficacy at the beginning of the course. All three of our models exhibited this outcome: how students interacted with one another in this active-learning environment predicted their physics self-efficacy regardless of how they scored on the self-efficacy survey at the beginning of the course. Our estimates suggest that a one standard deviation increase (or decrease) in a student's Pagerank would correspond to a 0.17 standard deviation change in a student's post physics self-efficacy, as seen in our best fit model (i.e. Model C). This association posits a direct link between students' social positioning in an active-learning classroom network and the confidence students develop within physics classroom contexts. Indeed, both the number of interactions that students have, as well as the kinds of people students interact with, matters for self-efficacy development in the subject area. This was not surprising given the particular emphasis that MI places on student interactions, and the theoretical relationship between interactions and self-efficacy development, which is particularly germane to vicarious learning and verbal persuasion – two socially construed sources of self-efficacy (Bandura, 1997; Sawtelle et al., 2012). Course developers should note the positive effect of student-centric and active-learning courses that nurture student engagement with a plurality of their classmates.

Although PageRank centrality in the classroom network did not directly predict physics interest at the end of the course, we tested whether a mediated relationship exists between these two variables via self-efficacy. We based this hypothesis on prior research indicating a directional link from self-efficacy to interest (Lent et al., 2008; Sheu et al., 2010; Smith & Fouad, 1999). Models with pathways going in both directions yielded similar fit measures (i.e. model A: post self-efficacy → post interest; model B: post interest → post self-efficacy), which hindered our ability to recognise self-efficacy as a mediator of PageRank centrality and students' interest. Nevertheless, our test of model C confirmed that students' incoming interest predicted their outgoing self-efficacy (not the other way around). This confirmed that self-efficacy does not mediate centrality and interest, and highlighted how in this context students who begin the course with high interest in physics are more likely to leave the course with higher physics self-efficacy. As posited by Nauta et al. (2002), this outcome suggests reconsidering the role of self-efficacy interventions to include a focus on interest development. Non-physics majors who come into the course with high physics anxiety and low interest would benefit from exposure that targets interest development.

Our models confirmed that students with high physics interest were more likely to develop high physics self-efficacy. From a theoretical perspective, both Bandura (1997) and Lent et al. (1994) propose that interest plays an important role in self-efficacy development during events where participants find themselves in novel situations or environments. When participants find themselves in new circumstances, interest helps to drive engagement in unfamiliar tasks. The successful or unsuccessful completion of those tasks, as well as the encouragement or emotions experienced during those tasks, could in turn shape participants' self-efficacy. Introductory physics courses often represent students' first sustained exposure to physics content, particularly for those who did not take physics in high school. We also know that students who participate in reformed or active-learning undergraduate physics courses may experience additional discomfort due to the lack of familiarity with course structure. When placed in environments where the rules for success are different and where activities require students to do more than just memorise and retain facts, students may develop negative beliefs that affect their participation (Vil-lasenor & Etkina, 2007). These negative perspectives can be exacerbated or meliorated by grading practices, instructor attitudes, and instructional norms (Turpen & Finkelstein, 2010). In this context, interest in physics as a subject helps to maintain (or hinder) student engagement in tasks that lead to self-efficacy development.

The extent to which our research can inform the development of active-learning curricula is limited by our design. While we saw decreases in self-efficacy and interest, confirmed pathways related to the development of those constructs, and situated class-room social networks as part of those pathways, we cannot directly attribute our outcomes to any particular aspect of the MI curriculum (e.g. small group activities, class discussions, model building, group assessments). Moreover, the context of our study is not easily replicated, occurring in an HSI where great cultural diversity exists even within groups who identify as Hispanic. This university further differs from others across the nation in size – among the largest in the country. These qualities inspire caution in the generalisations we make about our results. MI, too, as an active-learning introductory physics curriculum, differs in some ways from others implemented elsewhere. Still, our look at data across three different courses, two different instructors, and two separate academic years make us confident about the existence of these effects in our local community, and their possible existence in similar, though not necessarily identical, contexts. The work of others referenced in this paper, who have shown the broad applicability and effects of self-efficacy, interest, and active-learning pedagogies across a variety of cultures and contexts, suggests to us that the likelihood of finding these results in other university settings is not small.

## Acknowledgements

Much of this manuscript is based on the first author's Ph.D. thesis whose development was supported by the co-authors (Dou, 2017). We would like to thank the discipline-based education researchers associated with the STEM Transformation Institute led by Dr. Laird Kramer at Florida International University. Their work has both directly and indirectly contributed to this research. This work was funded by the National Science Foundation grant no. PHY 1344247. The authors declare no conflicts of interest, financial or otherwise, that may be perceived as influencing the authors' objectivity.

## Disclosure statement

No potential conflict of interest was reported by the authors.

## Funding

This work was funded by the National Science Foundation grant no. PHY 1344247.

## ORCID

Remy Dou http://orcid.org/0000-0001-8419-265X
Eric Brewe http://orcid.org/0000-0003-3480-7040
Justyna P. Zwolak http://orcid.org/0000-0002-2286-3208

Understanding the Development of Self-Efficacy

# Supplementary Material

# Social Network Survey

Panther ID:__________________                    Name:__________________

# Networks Survey – Physics

*Thank you for taking this survey. By completing it you contribute to research on the impact of social interactions in physics courses. We are interested in how networks form in and outside of classes. Please answer honestly and to the best of your knowledge.*

*Your answers will be kept anonymous and will not affect your success in this course or your classmates' success. Also, please note that students that you list will not know that you listed them in this survey and you will not know if anyone listed you.*

*For your convenience, we provided the names of all your classmates and instructors below. If you are not exactly sure of a name, choose your best guess. If you don't see a name on the list, you can still write it in the table.*

***Make sure to put your name and Panther ID on both pages!!!***

# Understanding the Development of Self-Efficacy

16. Name 1
11. Name 2
37. Name 3

-- Instructors --
85. name Instructor (PROF)
86. name LA (LA)

Understanding the Development of Self-Efficacy

Panther ID:_________________   Name:_________________

Question 1: Please choose from the presented list people from your physics class that you had a meaningful interaction with **in class** this week, even if you were not the main person speaking or contributing. You may include names of students outside of the group you usually work with. You don't have to fill in all columns. You may use the name of a student or their corresponding number.

| I had a meaningful interaction with these people this week in class... | | |
|---|---|---|
| … one time. | … more than one time but NOT every day. | … every day. |
| | | |

Question 2: Throughout this week, did you work with anyone on physics-related material **outside of class**, either in person or virtually (using, e.g., WhatsApp, Google Chat, etc.)? If so, whom did you work with? (*Please provide their FIRST and LAST name. If you worked with someone not from your class, provide their major if possible.*) If not, please write NONE. If you worked with someone from a school-related group you participate in, please name that group. Also, please name instructors or TAs if you met with one or more out-of-class.

Table S1. Standardized item path coefficients for structural equation models.

| | Model A | | Model B | | Model C | |
|---|---|---|---|---|---|---|
| | Pre | Post | Pre | Post | Pre | Post |
| **Physics interest** | | | | | | |
| Mechanics | 0.71 | 0.65 | 0.71 | 0.65 | 0.71 | 0.65 |
| Optics/Waves | 0.77 | 0.81 | 0.77 | 0.81 | 0.77 | 0.81 |
| Electromagnetism | 0.79 | 0.88 | 0.79 | 0.88 | 0.79 | 0.88 |
| Relativity/Modern Physics | 0.78 | 0.85 | 0.78 | 0.86 | 0.77 | 0.85 |
| History and people of physics | 0.47 | 0.51 | 0.47 | 0.51 | 0.46 | 0.51 |
| Current topics in physics | 0.66 | 0.75 | 0.66 | 0.75 | 0.67 | 0.75 |
| **General Science Interest** | | | | | | |
| Understanding natural phenomena | 0.68 | 0.68 | 0.68 | 0.68 | 0.69 | 0.68 |
| Understanding science in everyday life | 0.70 | 0.78 | 0.69 | 0.78 | 0.71 | 0.78 |
| Explaining things with facts | 0.70 | 0.79 | 0.71 | 0.80 | 0.70 | 0.79 |
| Telling others about science concepts | 0.77 | 0.76 | 0.77 | 0.76 | 0.76 | 0.76 |
| Making scientific observations | 0.71 | 0.76 | 0.71 | 0.75 | 0.70 | 0.76 |
| Wanting to know more science | 0.73 | 0.83 | 0.73 | 0.82 | 0.73 | 0.83 |
| **Physics Self-efficacy** | | | | | | |
| Item I | 0.52 | 0.61 | 0.53 | 0.61 | 0.53 | 0.61 |
| Item II | 0.63 | 0.57 | 0.63 | 0.57 | 0.63 | 0.57 |
| Item III | 0.59 | 0.63 | 0.59 | 0.64 | 0.59 | 0.63 |
| Item IV | 0.62 | 0.68 | 0.62 | 0.68 | 0.63 | 0.68 |
| Item VI | 0.37 | 0.49 | 0.37 | 0.49 | 0.37 | 0.49 |
| Item VII | 0.62 | 0.74 | 0.62 | 0.74 | 0.63 | 0.74 |
| Item VIII | 0.62 | 0.77 | 0.62 | 0.77 | 0.62 | 0.77 |
| Item X | 0.51 | 0.51 | 0.51 | 0.51 | 0.51 | 0.51 |
| Item XI | 0.53 | 0.71 | 0.53 | 0.71 | 0.53 | 0.70 |
| Item XIV | 0.69 | 0.74 | 0.69 | 0.74 | 0.69 | 0.74 |
| Item XV | 0.52 | 0.62 | 0.52 | 0.62 | 0.52 | 0.62 |
| Item XVII | 0.67 | 0.67 | 0.67 | 0.67 | 0.67 | 0.67 |
| Item XVIII | 0.60 | 0.10 | 0.60 | 0.11 | 0.59 | 0.12 |
| Item XIX | 0.49 | 0.49 | 0.49 | 0.50 | 0.50 | 0.49 |
| Item XX | 0.55 | 0.58 | 0.55 | 0.58 | 0.55 | 0.58 |
| Item XXI | 0.67 | 0.76 | 0.67 | 0.77 | 0.67 | 0.77 |
| Item XXII | 0.52 | 0.43 | 0.52 | 0.43 | 0.52 | 0.42 |

| Item | | | | | | |
|---|---|---|---|---|---|---|
| Item XXVI | 0.51 | 0.41 | 0.51 | 0.42 | 0.51 | 0.41 |
| Item XXVII | 0.62 | 0.52 | 0.62 | 0.53 | 0.62 | 0.52 |
| Item XXIX | 0.48 | 0.62 | 0.48 | 0.62 | 0.48 | 0.62 |
| Item XXX | 0.45 | 0.43 | 0.46 | 0.43 | 0.46 | 0.42 |
| Item XXXI | 0.61 | 0.63 | 0.61 | 0.63 | 0.61 | 0.62 |
| Item XXXII | 0.45 | 0.42 | 0.45 | 0.42 | 0.45 | 0.41 |